\documentclass[showpacs,preprintnumbers,10pt,twocolumn]{revtex4-1}
\usepackage{dcolumn}
\usepackage{appendix}
\usepackage{bm}
\usepackage{graphicx}
\usepackage{epstopdf}
\usepackage{amsmath,amssymb,amsthm}
\usepackage[mathscr]{eucal}
\usepackage{changes}

\begin{document}
\title{Dissipatively stabilized quantum sensor based on indirect nuclear-nuclear interactions}

\author{Q. Chen$^{\dag}$}
\author{I. Schwarz$^{\dag}$}
\author{M.B. Plenio$^{}$ }
\affiliation{ Institut f\"{u}r Theoretische Physik, Albert-Einstein-Allee 11, Universit\"{a}t Ulm, 89069 Ulm, Germany\\
$^{\dag}$ These authors contributed equally to this work}

\begin{abstract}
We propose to use a dissipatively engineered nitrogen vacancy (NV) center as a mediator of interaction between
two nuclear spins that are protected from decoherence and relaxation of the NV. Under ambient conditions this scheme
achieves highly selective high-fidelity quantum gates between nuclear spins in a quantum register even at large NV-nuclear
distances. Importantly, this method allows for the use of nuclear spins as a sensor rather than a memory, while the
NV spin acts as an ancillary system for the initialization and read out of the sensor. The immunity to the decoherence
and relaxation of the NV center leads to a tunable sharp frequency filter while allowing at the same time the continuous
collection of the signal to achieve simultaneously high spectral selectivity and high signal-to-noise ratio (SNR).
\end{abstract}
\maketitle

\emph{Introduction ---}
The Nitrogen vacancy (NV) center is attracting increasing attention due to the possibility for creating hybrid
quantum registers of nuclear spins controlled by the electron spin of the NV center \cite{taminiau2014universal,
waldherr2014quantum,PhysRevLett.117.130502,bermudez2011electron} and due to its applications in nanoscale sensing \cite{waldherr2012high,taminiau2012detection,kolkowitz2012sensing,zhao2012sensing,kaufmann2013detection,ermakova2013detection,
mamin2013nanoscale,staudacher2013nuclear,grinolds2014subnanometre,muller2014nuclear,sushkov2014magnetic,shi2015single,
trifunovic2015high,lovchinsky2016nuclear,laraoui2013high,ajoy2015atomic,greiner2015indirect,wang2016delayed,zaiser2016enhancing,pfender2016nonvolatile,rosskopf2016quantum,SchmittGS2016,ajoy2016dc}. However, for both quantum register and sensing applications, there are several outstanding challenges caused by 
the relaxation and decoherence processes of the NV center electron spin as these limit quantum gate fidelities 
on nuclear registers as well as spectral resolution, selectivity and signal to noise ratio in sensing applications. 
While the addition of the nuclear ancilla as long-lived memory of the NV sensor can improve spectral resolution,
by extending the time interval between interrogations, e.g. in correlation spectroscopy, this comes at a price of 
signal to noise ratio and remains limited by the relaxation time of the NV center.

\begin{figure}
\center
\includegraphics[width=3.3in]{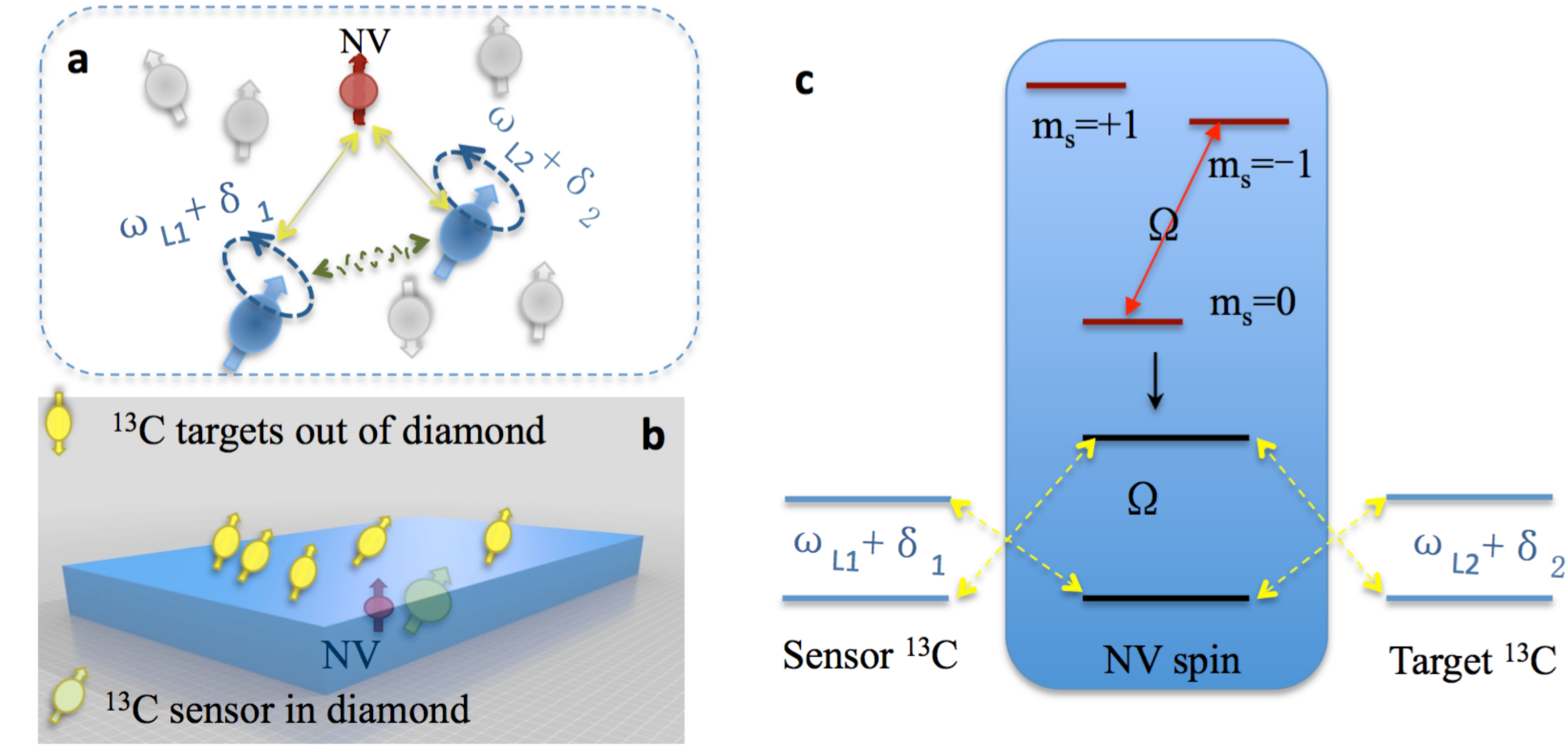}
\caption{(a) Illustration of the basic principle. An NV center mediates the coupling between
two nuclear spins, while itself becoming decoupled from the dynamics, enabling long coherent interactions,
e.g. for high fidelity selective quantum gates. Dissipation-assisted sensing set-up
composed of a nuclear quantum sensor (a single $^{13}$C spin (green) $\sim1$ nm removed from
the associated NV center (red)) and an NV implanted 3-4 nm below the diamond surface
where the target spins reside (yellow). (b) Energy level diagram. The Rabi frequency
$\Omega$ of the MW drive is far off resonance from the Larmor frequency of the target nucleus which is
on resonance with the sensor. }
\label{setup}
\end{figure}

Here we address directly the challenge of the NV center decoherence and relaxation by using the NV center
as a mediator to couple two nuclear spins, while eliminating the NV center and the effect
of its decoherence and relaxation from the dynamics (Fig.~\ref{setup}a). Thus, coherent evolution between
nuclear spins is achieved which is no longer limited by the NV decoherence or even relaxation. The key idea
is the use of the substantial second order coupling between the nuclear spins obtained through a strongly
detuned NV center which is periodically reinitialized by a dissipative process \cite{BermudezSP2013}.
The detuning and periodical reinitialization of the NV center decouple it from the dynamics and its effect
on the system can be modeled by an effective but weak dissipation process. Thus, high fidelity of selective
quantum gates between the nuclear spins, as well as an improved sensing setup, are made possible even at
ambient condition.

We demonstrate that the NV-mediated interaction between nuclear spins yields a dissipation-enhanced
quantum sensor by considering a basic setup including a single $^{13}$C nuclear spin sensor nearby the
shallow NV center to sense a target $^{13}$C spins on the diamond surface, as shown in Fig. \ref{setup}b.
Contrary to previous applications of nuclear spins acting as ancillary memory \cite{laraoui2013high,ajoy2015atomic,greiner2015indirect,wang2016delayed,zaiser2016enhancing,pfender2016nonvolatile,rosskopf2016quantum,SchmittGS2016,ajoy2016dc}, the nuclear spin in our set-up serves directly
as a sensor, while the NV spin is an ancillary mediator of interaction, and at beginning and end
serves as a means for initialization and read out of the sensor spin. The effective coupling between the two
nuclear spins is controllable, which provides a tunable sharp frequency filter for achieving very high
spectral resolution that is not limited by the NV decoherence and relaxation. Additionally, the
signal is continuously accumulated by the nuclear sensor, with no dead-time, hence maintaining sensitivity by increasing
the SNR. Furthermore our scheme does not merely measure Larmor frequency \cite{waldherr2012high} but retains
the option of individually addressing the nuclear spins, useful for 2D spectroscopy \cite{ajoy2015atomic,KostCP2015,ScheuerSK+2015} and quantum registers.

\emph{Effective master equation ---}
The basic setting we consider here includes a single NV center spin and two $^{13}$C nuclear spins.
Applying a weak magnetic field $\vec{B}_0$ lifts the degeneracy of $|m_s = -1\rangle$ and $|m_s =
+1\rangle$ to allow for selective continuous microwave (MW) driving field of one
specific electronic transition. Working in a frame that is rotating with the MW frequency,
resonant with the $|m_s=0\rangle \leftrightarrow |m_s=-1\rangle$ transition, the effective Hamiltonian
of the NV spin is $H_{NV}=\Omega\sigma_z$. Here, $\Omega$ is the Rabi frequency of the MW
drive and $\{|+_x\rangle=\frac{1}{\sqrt{2}}(|0\rangle+|-1\rangle),|-_x\rangle = \frac{1}{\sqrt{2}}
(|0\rangle-|-1\rangle)\}$ are the MW dressed eigenstates. Our goal is to use such MW dressed
NV center spin as a mediator for indirect coherent interactions between two nuclear spins. The interaction
between the NV and a nuclear spin is given by
\begin{eqnarray}
    H_{intf}&=& S_z\vec{A}_i\cdot\vec{I}_i= S_z(a_{\parallel i}I^z_{i}+a_{\perp i}I^{\hat{x}}_{i}),
\end{eqnarray}
where $\vec{A}_i$ is the hyperfine coupling vector and the direction of $\vec{I}_i$ is determined by
the magnetic field unit vector $\vec{b}(\theta,\phi)$. $S_q$ and $I^q_{i}$ are the NV electronic spin
and external nuclear spin operators respectively. $\vec{A}_i=(a_{\parallel_i}, a_{\perp_i})$ with
$a_{\parallel_i}$ and $a_{\perp_i}$ denoting the parallel and perpendicular coupling components to
the nuclear spin quantization axes $a_{\parallel i}=\vec{A}_i\cdot\vec{b}$ and $a_{\perp i} =
\sqrt{|\vec{A}_i|^2-a_{\perp i}^2}$.

We derive a Lindblad master equation to describe the dynamics of the whole system $\rho$, which is
described as $\frac{d}{dt}\rho=-i[H'_{tot},\rho]+\mathcal{D}[\rho]$ in which $\mathcal{D}[\rho] = \mathcal{D}_e[\rho]+\mathcal{D}_n[\rho]$, corresponding to the relaxation and dephasing of the electron
and nuclear spins, respectively. The effective total Hamiltonian of the three spins are thus rewritten
as (for details see SI \cite{SI})
\begin{eqnarray}%
    H'_{tot}&=&\Omega\sigma_z+\sum_{i=1}^2(\gamma_{ni}\vec{B}_0\vec{I}_i+\frac{\vec{A}_i\vec{I}_i}{2})
    +\sigma_x\vec{A}_i\vec{I}_i,
    \label{Htot}
\end{eqnarray}%
with $\gamma_{ni}$ the gyromagnetic ratio of the nuclear spin. The NV center is reinitialized
periodically (every $t_{re}$) to the state $|-_x\rangle$ of the dressed state basis, namely,
$\rho(N t_{re})\to [ \text{Tr}_e\rho(N t_{re})]\otimes|-_x\rangle\langle-_x|$, where $ \text{Tr}_e$ denotes the
partial trace over the electron spin and $N$ is an integer.

The reinitialization of the NV results in nuclear subsystem evolution in the vicinity of the NV quasi-steady
state, $\rho_{ss}^{NV}=p_+ |+_x\rangle\langle+_x|+p_- |-_x\rangle\langle-_x|$, in which $p_+ = 1-\frac{e^{-t_{re}/T_{1\rho}}}{2t_{re}/T_{1\rho}}$ when $t_{re}\le T_{1\rho}$ and $p_+ = \frac{1}{2}
-\frac{1-e^{-1}}{2t_{re}/T_{1\rho}}$ when $t_{re}> T_{1\rho}$, $p_-=1-p_+$. Notice that the effective
relaxation rate of the NV spin is determined by the lifetime $T_{1\rho}$ and the reinitialization
time $t_{re}$, $\Gamma_N=1/T_{1\rho} + 1/t_{re}$. We can now use the electron spin as a bus to
mediate the nuclear coupling \cite{bermudez2011electron,BermudezSP2013}, if the relevant energy scales
satisfy $\Omega, \gamma_{n1}B_0, |\Omega- \gamma_{n1}B_0|\gg |\vec{A}_i|$. The Schrieffer-Wolff
transformation for open systems \cite{cohen1992atom, schrieffer1966relation} then allows for the
derivation of an effective Liouvillian (see SI for details) \cite{SI} for the nuclear subsystem $\rho_n$
\begin{eqnarray}%
    \label{LindladN1}
    \frac{d}{dt}\rho_n&=&-i[H_{eff},\rho_n]+\mathcal{D}_{n}[\rho_n]+\mathcal{D}_{eff}[\rho_n].
\end{eqnarray}%
Here the coherent Hamiltonian is
\begin{eqnarray}%
    H_{eff} &\approx&\sum_{i=1,2}(\omega_{Li}+\delta_{i})I^{z}_i + pA_{wo}(I^+_1I^{-}_2+I^-_1I^{+}_2),
    \label{HWO}
\end{eqnarray}%
in which we have $p=p_+-p_-$, $\omega_{Li}=\gamma_{ni}B_0$ and
\begin{equation}
  \delta_{1} \approx\frac{a_{\parallel_i}}{2}-\frac{a_{\perp_i}^2}{16}(\frac{\Delta_{-i}}{\Delta_{-i}^2+(\frac{\Gamma_N}{2})^2}
    +\frac{\Delta_{+i}}{\Delta_{+i}^2+(\frac{\Gamma_N}{2})^2}),
    \label{delta}
\end{equation}
and
\begin{equation}
  A_{wo} \approx\sum_{i=1,2}\frac{a_{\perp_1}a_{\perp_2}}{32}(\frac{\Delta_{-i}}{\Delta_{-i}^2 +(\frac{\Gamma_N}{2})^2}+\frac{\Delta_{+i}}{\Delta_{+i}^2+(\frac{\Gamma_N}{2})^2})
    \label{Awo}
\end{equation}
with $\Delta_{\pm i}=\Omega\pm(\gamma_{ni}B_0 + \frac{a_{\parallel i}}{2})$. The virtual NV spin excitation
results in an indirect dissipation contribution to Eq. (\ref{LindladN1}),
\begin{eqnarray}%
    \mathcal{D}_{eff}[\rho_n]\approx\sum_{i,j=1,2}\Gamma^{eff}_{ij}\Big(I^{-}_i\rho_nI^+_j - \frac{1}{2}\{I^+_iI^{-}_j,\rho_n\}_+\Big),\nonumber
\end{eqnarray}%
with
$\Gamma^{eff}_{ij}=\sum_{j=1,2}\frac{a_{\perp_i}a_{\perp_j}}{32}(\frac{\Gamma_N}{\Delta_{+j}^2+(\frac{\Gamma_N}{2})^2}
    +\frac{\Gamma_N}{\Delta_{-i}^2+(\frac{\Gamma_N}{2})^2})$. Notice that the periodic reset of the NV
maintains a net NV polarization ($p>0$), which is critical to our scheme. The NV center provides
two channels for virtual electron spin flips, via $|-_x\rangle$ and $|+_x\rangle$ \cite{SI}, that
mediate interaction between the nuclei. As the two channels produce effective nuclear couplings
with opposing sign, in the absence of periodic reinitialization, the mixing of the two channels would
induce no net coherent evolution ($p=0$) for an evolution time that is longer than the NV lifetime
$T_{1\rho}$ while a net interaction is achieved in the presence of periodic reinitialization ($p>0$).

\begin{figure}
\center
\includegraphics[width=1.6in]{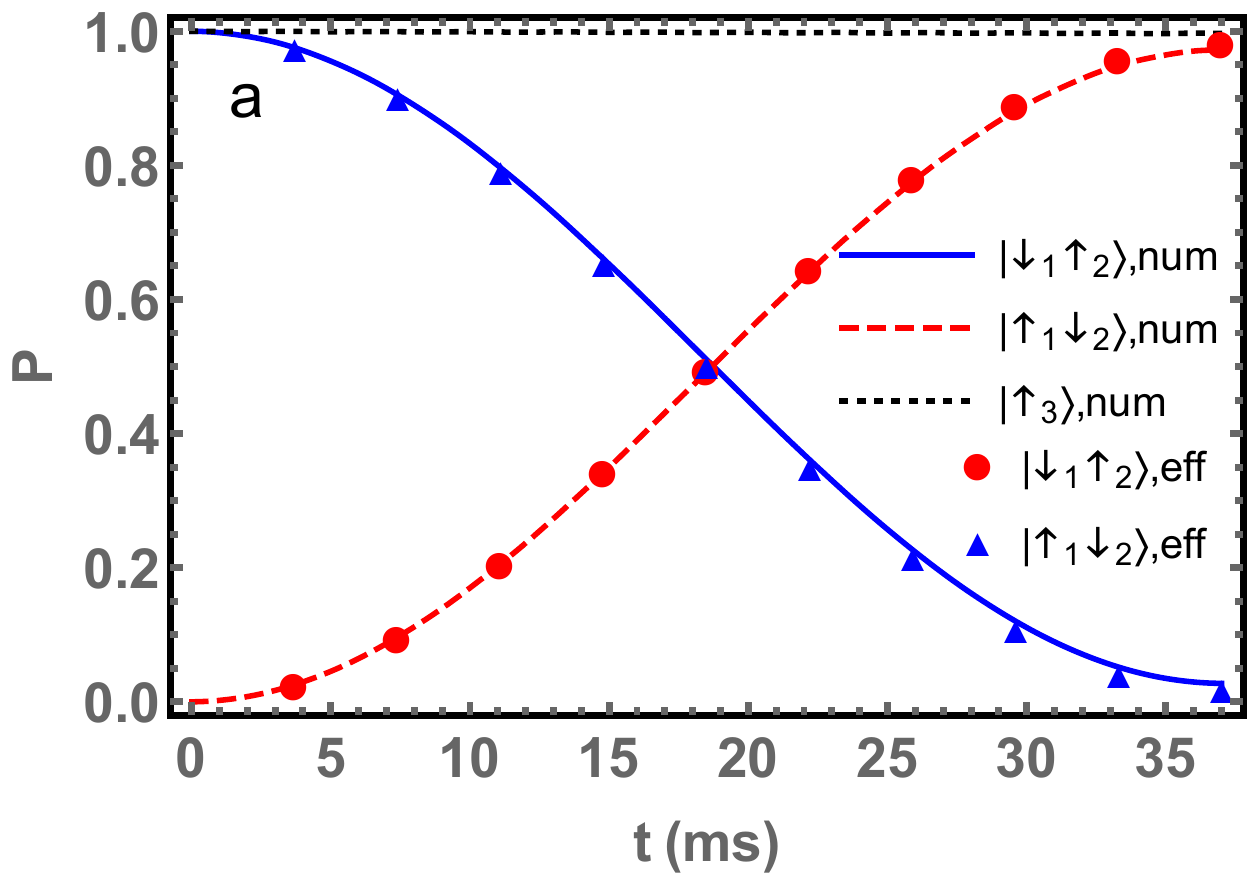}
\includegraphics[width=1.6in]{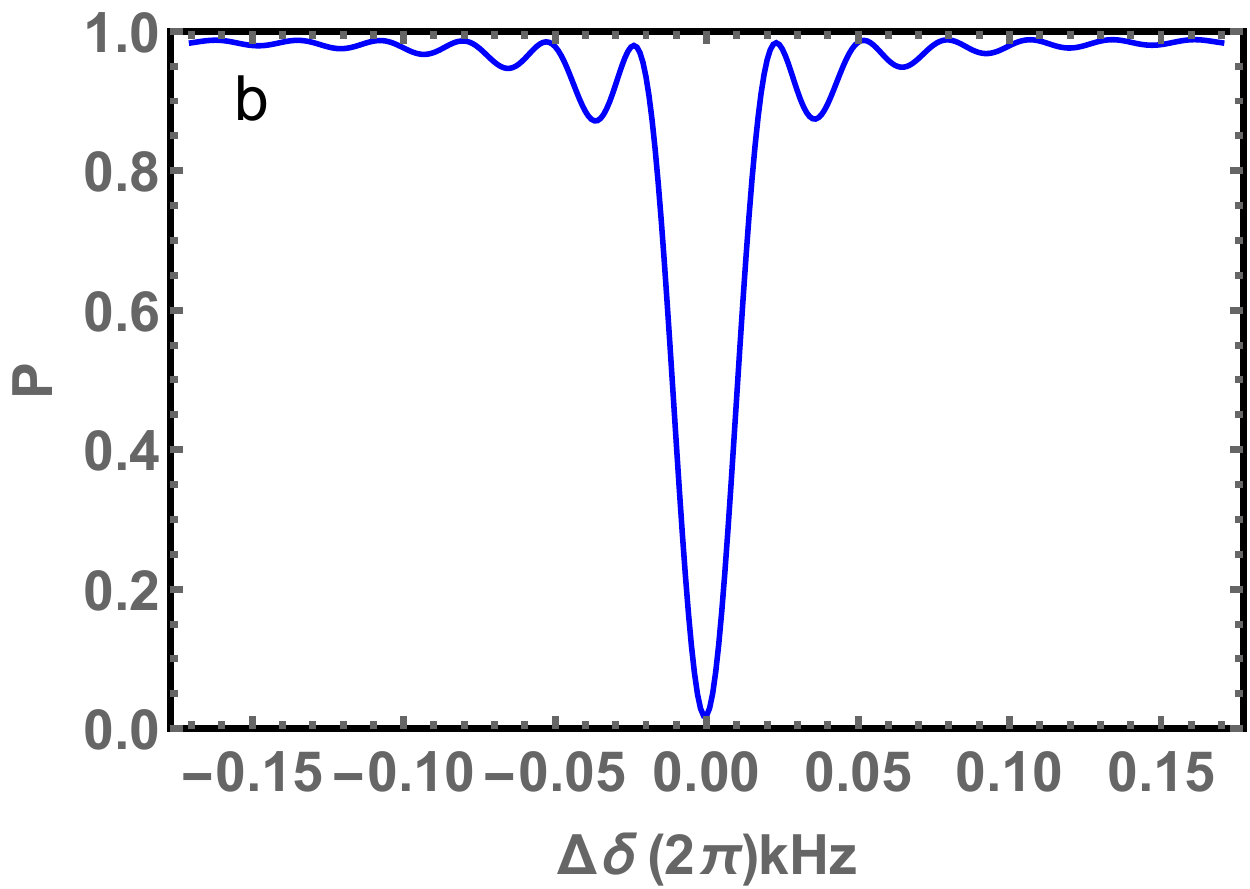}
\caption{ (a) Comparison of the effective (eff) dynamics under master equation (\ref{LindladN1}) and the exact
numerical simulation (num) by using full Hamiltonian equation (\ref{Htot}) and NV resets applied
every $T_{1\rho}=t_{re}=1$ ms. The solid blue (dashed red) curve denotes the probability of measuring $|\downarrow,\uparrow\rangle$($|\uparrow,\downarrow\rangle$) as a function of time $t$. Choosing
$\Omega=(2\pi) 300$ kHz, $\gamma_{ni}B_0=(2\pi) 200$ kHz for the NV center, we achieve a near
perfect coherent flip-flop between nuclear spins $1$ and $2$ with $(a_{\parallel_1}, a_{\perp_1}) =
(2\pi)(1.99,2.01)$ kHz, $(a_{\parallel_2}, a_{\perp_2})=(2\pi)(2.00,5.01)$ kHz that are coupled
via the NV spin, while the nuclear spin 3 with $(a_{\parallel_3}, a_{\perp_3})=(2\pi)(2.30,2.01)$ kHz
is not affected. (b) The parameters are the same as in (a), with the blue line denoting the
occupation probability of the state $|\downarrow,\uparrow\rangle$ for the evolution time $T=37$ ms. The
high selectivity is demonstrated by assuming a shift of the detuning of the second spin from the
resonance condition, with $\Delta \delta=\delta_i-\delta_2$.}
\label{signal}
\end{figure}
%
%At $19$ ms this would represent a $\sqrt{SWAP}$ gate, universal for quantum computation, with \mbp{$\sim 0.99$} fidelity.
%
\begin{figure*}[t]
\center
\includegraphics[width=2.2in]{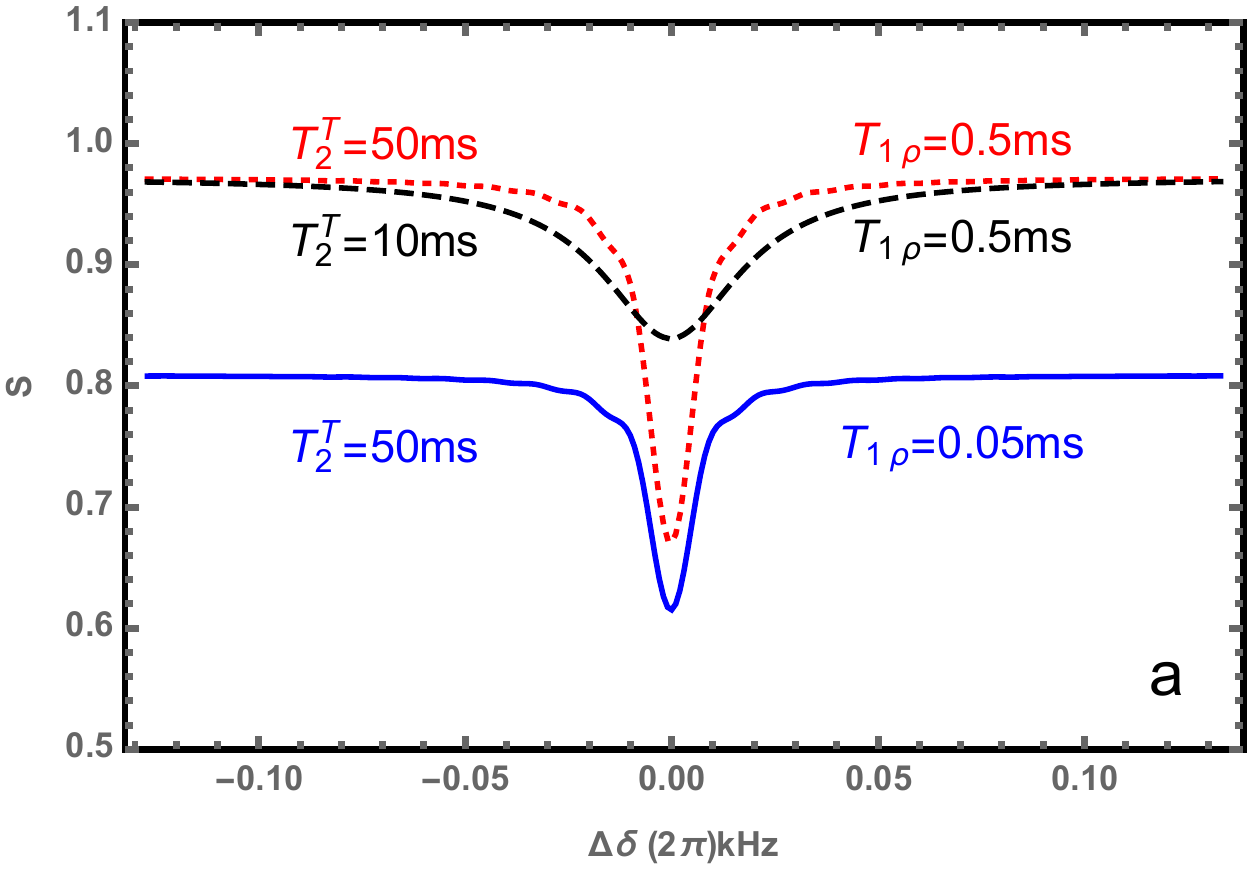}
\includegraphics[width=2.2in]{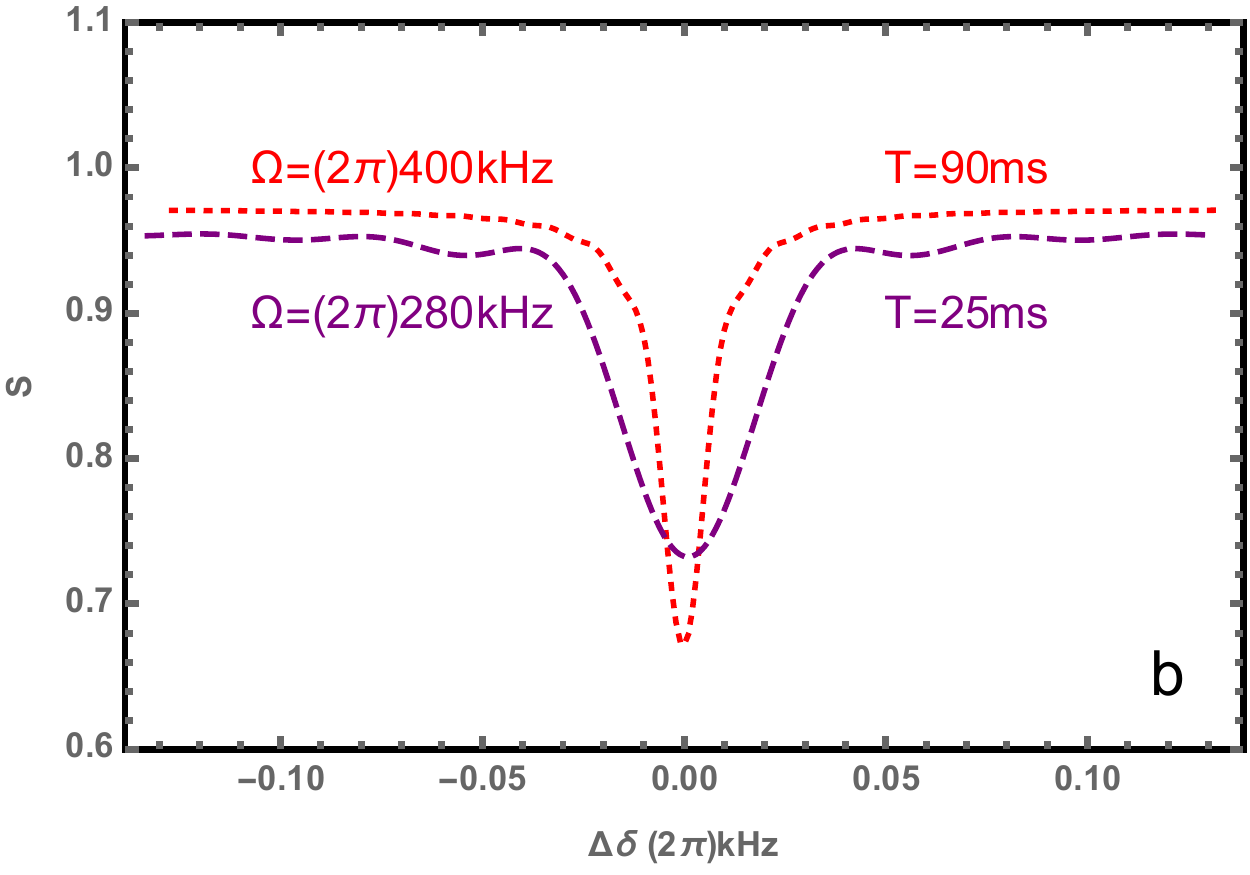}
\includegraphics[width=2.2in]{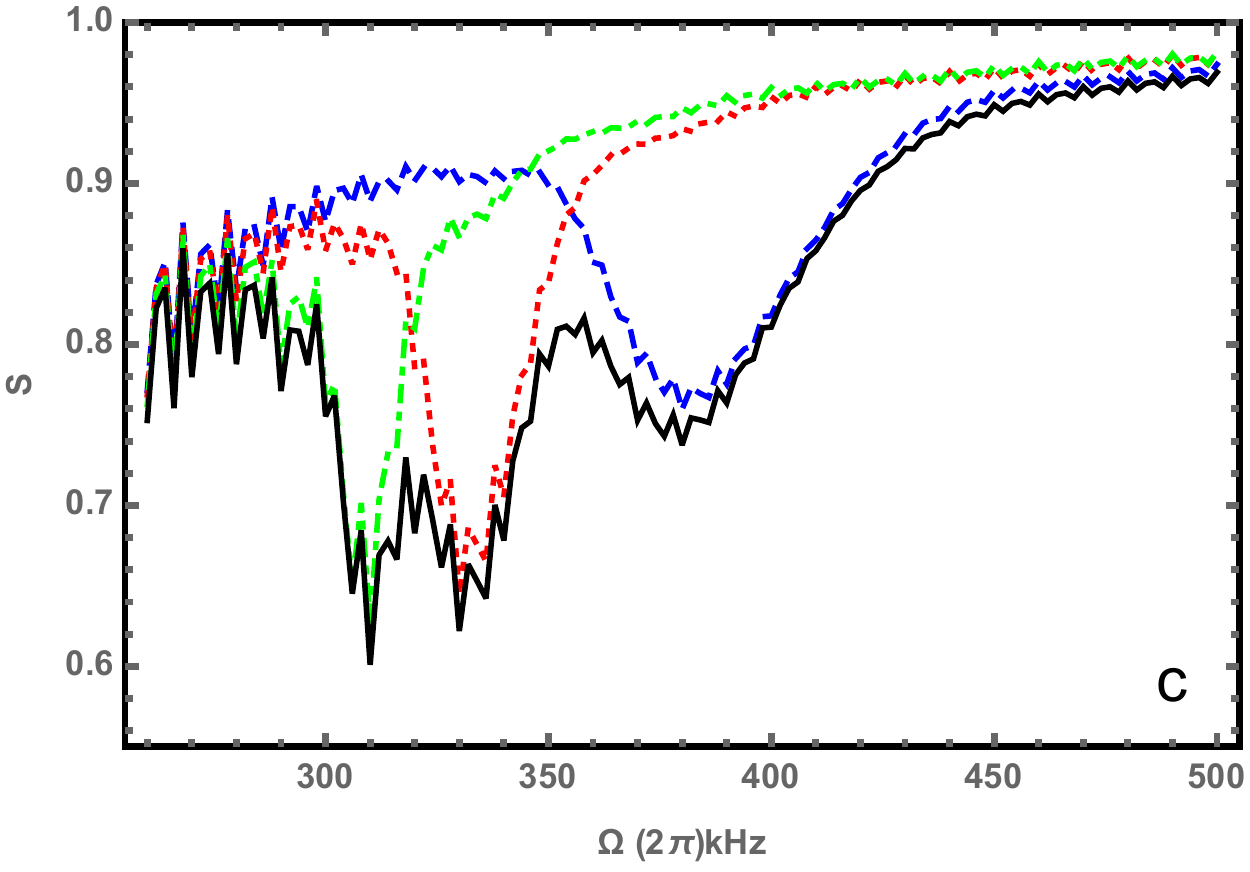}
\caption{(a), The effect of the NV $T_{1\rho}$ and the target spin decoherence time $T^T_2$
on the detection bandwidth. The sensor signal $S$ (probability that the sensor spin is found
in its initial state) is calculated by exact simulation for a total run time of $T=90$ ms and three
combinations of $T_{1\rho}$, $T^T_2$. The NV spin is reinitialized every $t_{re}=T_{1\rho}$
and we chose $\omega_{L1} = (2\pi) 200$ kHz, $\Omega=(2\pi) 400$ kHz and couplings $a_{\perp_1}
=(2\pi)10$ kHz for the sensor nucleus and $a_{\perp_2}=(2\pi)1$ kHz for the target nucleus.
It can be seen that only the sensor $T^T_2$ has an effect on the linewidth, while the NV $T_1^{\rho}$
only affects the signal strength. (b), The tunable NV detuning $\Delta_{\pm i}$ as a frequency
filter - the red dotted line has identical parameters to that in (a), the purple dashed line with
different MW Rabi frequency results in smaller detuning which leads to increased
linewidth due to increased $A_{wo}$ and $\Gamma_{ij}^{eff}$. (c), Assuming three $^{13}$C nuclear spins
in a Valine molecule. The NV spin is at the origin, the sensor at $[-0.601, 0.676, -0.692]$nm,
and targets at $[-1.260, -1.451, 2.904]$nm, $[-1.260, -1.317, 3.135]$nm and $[-1.260, -1.317, 2.673]$nm,
with the magnetic field direction $[44.7^\circ, 52.0^\circ]$ and $T=60$ ms. The dashed lines show
individual contribution of the three $^{13}C$ spins with different coupling components to the
NV spin, while the black solid line represents the total signal of the three spins. The internuclear
coupling has been suppressed by means of dynamical decoupling. The small amplitude oscillations 
on the observed signal are due to off-resonant contributions, which can be suppressed by choosing 
larger detunings $\Delta_{\pm i}$, i.e. larger Rabi frequency $\Omega$.}
\label{spectra}
\end{figure*}

\emph{Quantum gate implementation ---} The effective coupling term in Eq. (\ref{HWO}) gives rise to
coherent flip-flop processes between the two nuclear spins, which achieves very high fidelity under the
condition $pA_{wo}\gg\Gamma^{eff}$. As the effective relaxation rate $\Gamma^{eff}$ is inversely
proportional to the square of the detuning $\Delta_{\pm i}$, we can adjust $\Delta_{\pm i}$ and the
NV reset time, i.e., a large detuning $\Delta_{\pm i}\gg\Gamma_N$ and $t_{re}\sim T_{1\rho}$ to
minimize this induced dissipation. Thus, it is possible to have coherent internuclear evolution well
beyond the life time of NV spin. In Fig. \ref{signal}a, we initialize two nuclear spins in state
$|\downarrow_1\uparrow_2\rangle$ and a third one in state $|\uparrow_3\rangle$. The coherent flip-flop
interaction between $|\downarrow_1\uparrow_2 \rangle \leftrightarrow |\uparrow_1\downarrow_2\rangle$
induces an nuclear XX gate between the nuclear spins weakly coupled to NV spin with $T_{1\rho}=1$ ms.
We compute the average fidelity to find $0.994$ \cite{nielsen2002simple}, significantly higher than the
fidelity ($<0.66$) of nuclear-nuclear gates achieved so far with NV centers by using $4$ electron-nuclear spin quantum
gates (if each fidelity $<0.90$ \cite{taminiau2014universal}). For the $^{13}C$ nuclear spin bath
surrounding an NV spin, in principle, one could select two weakly nuclear spins to implement a near
perfect quantum gate at room temperature. For quantum registers that are composed of several
nuclei, it is important that quantum gates between two spins do not affect the other qubits in
the register. As shown in Fig. 2a, our scheme can achieve very high selectivity. Even a
$0.3$ kHz deviation of the parallel component coupling from the resonance leaves a third nuclear
spin unaffected by the quantum gate, see Fig. 2b for the high specificity of the interaction.
The resulting spatial resolution will be further discussed later.

\emph{Detection of a single nucleus ---} The coupling between the nuclear spins mediated by
a dissipative NV can also be used for sensing an isolated single nucleus outside of the diamond.
Our scheme includes three steps: (1), Polarize the $^{13}C$ nuclear sensor spin by using the
NV center electron spin. (2), Achieve polarisation transfer via a frequency selective flip-flop
process between the two nuclear spins that is mediated by a dissipative NV center. (3), Use
NV center to read out the polarization leakage of the nuclear $^{13}C$ spin sensor, which indicates
the existence of a distant nuclear spin at the selected frequency. The electron spin
of the NV center can be optically initialized and read out by using laser illumination. The nearby
nuclear $^{13}$C sensor spin can also be well-controlled by a SWAP gate between NV and nuclear spin,
i.e., it can be polarized and its position can be detected. A high-fidelity SWAP operation can be
constructed by using radio-frquency (rf) and MW pluses \cite{wang2016delayed,jiang2009repetitive} or
via the use of decoherence protected gates \cite{cappellaro2009coherence}.

Therefore if the sensor $^{13}C$ spin is polarized to the ground state $| \downarrow_1\rangle$, we
can detect the target $^{13}C$ nuclear spin by sensing the polarization leakage of the sensor. After
time $t$, we measure the probability that the sensor $^{13}C$  spin remains in the state $| \downarrow_1
\rangle$ which depends on the effective internuclear coupling strength $A_{wo}$ as follows
\begin{eqnarray}%
    S = 1 - \frac{(pA_{wo})^2\sin^2(\frac{t}{2}\sqrt{(pA_{wo})^2 + (\delta_{1}-\delta_{2})^2})}{2[(pA_{wo})^2 +(\delta_{1}-\delta_{2})^2]}.
    \label{S}
\end{eqnarray}%
It is easy to find the resonance condition $\delta_{1} = \delta_{2}$.
In our setup, the coupling of the nuclear sensor (which is labeled spin 1 and whose position can
be determined precisely) to the NV is far larger than that of the nuclear target to the NV. In order
to match the resonance condition to the target nuclear spin, two main degrees of freedom are available-
we can adjust the direction of the magnetic field which gives a coarse tuning, and the Rabi frequency
of the microwave drive on the NV spin, which allows a fine-tuning to match the resonance condition
\cite{SI}. Note that the Rabi frequency has only second order contributions to the resonance condition,
i.e., $\frac{a_{\perp_i}^2}{16\Delta_{\pm i}}$, thus our scheme is not sensitive to small deviations or
fluctuations of the Rabi frequency. Once the resonance condition is achieved, the signal dips around the
time $t=\pi/4pA_{wo}$ marking the presence of nuclear spin.

\emph{Enhanced spatial resolution ---} When using the NV as a sensor by observing the polarization
leakage from the NV center to the nuclear spins \cite{cai2013diamond}, there exist two major limitations
for the frequency resolution: that the NV $T^\rho_1$ relaxation has to be longer than the NV-nuclear
flip-flop time and the requirement that the hyperfine transverse component is small ($|a_{\parallel_i} - a_{\parallel_j}|/2> a_{\perp_{i/j}}$). Our scheme solves both of these limitations. First, in our
scheme the influence of NV decoherence ($T_2$) and relaxation ($T_1$) on the dynamics can be reduced at
will below any desired level by increasing detuning $\delta_{\pm i}$ and NV reset rate. Therefore the resolution is limited
only by the target spin decoherence $T_2^T$ (see Fig. \ref{spectra}a). Second, the broadening due to $a_{\perp_{i}}$
now depends quadratically on $a_{\perp_{i}}$ (see Eq. (\ref{delta})) and is essentially proportional to
the effective coupling $pA_{wo}$ (see Eq. (\ref{Awo})), which is at least one order of magnitude smaller
than $a_{\perp_{i/j}}$. Thus the frequency resolution is improved accordingly as we now need to require
$|\delta^w_i-\delta^w_j| \simeq |a_{\parallel_i}-a_{\parallel_j}|/2 > pA_{wo}$. Moreover, increasing the
detuning $\Delta_{\pm i}$ of the NV induces a smaller effective coupling $pA_{wo}$ enhancing the frequency resolution (see Fig.
\ref{spectra}b), which also gives a tunable frequency filter. For optimal resolution, one would fix the
NV detuning $\Delta_{\pm i}$ to balance the two limitations, i.e.  $\frac{1}{pA_{wo}} \lesssim T^T_2$. For
example, for $T_{1\rho}=0.05$ ms and $T^T_2=50$ ms, the resolution could be improved a thousand-fold
compared to direct sensing via the NV center spin.

Recently the use of a nuclear spin as ancillary memory to enhance the resolution of NV detection has 
been proposed for correlation spectroscopy~\cite{laraoui2013high,zaiser2016enhancing,pfender2016nonvolatile}. 
However, the enhanced spectral resolution is offset by other disadvantages, not present in our proposed 
scheme, which are especially limiting for the regime of interest in NV sensing, i.e. an NV $\sim3-5$ 
nm from the surface, attempting to sense single nuclear spins ($a_{\perp}\approx 1$ kHz). In correlation 
spectroscopy the interaction time with the target nuclei is limited by the $T_2^{NV}$ of the NV, while 
the longest time interval between interrogations, and hence the spectral resolution, is limited by the 
$T_2^{ancilla}$ time of the nuclear ancilla. Therefore, the SNR for a given unit time is reduced by a 
factor of $T^{NV}_2/T^{ancilla}_2 < 0.01$ due to the relatively long "dead-time" in between measurements. 
Ramsey spectroscopy \cite{waldherr2012high} by using an NV-center is limited due to the required initialization 
and the readout of the nuclear spin. The efficient initialization of the nuclear spin requires $a_{\perp} 
T^{NV}_2\ll 1$ and spectrally selective addressing is limited by $T^{NV}_2$. In correlation spectroscopy
the SNR is reduced due to long time interval between measurements. In contrast, in our set-up it is the 
nuclear spin which serves directly as a sensor which is not limited by the decoherence and relaxation of 
the NV spin. One can estimate the sensitivity of our scheme as follows: suppose $t_{re}=T_{1\rho}$, as both 
$\sqrt{1/\Gamma_{eff}}$ and $pA_{wo}$ are linearly dependent on the detunings $\Delta_{\pm i}$, the 
sensitivity per unit time of our scheme is proportional to $\frac{pA_{wo}}{\sqrt{1/\Gamma_{eff}}}\sim \frac{a_{\perp_2}/4}{\sqrt{T_{1\rho}}}$, 
which is independent of $\Delta_{\pm i}$. This is 
of the same order as the sensitivity achieved by direct detection of a nucleus by polarisation leakage 
of the NV \cite{cai2013diamond} while obtaining enhanced spectral resolution. Therefore, our scheme 
achieves the enhanced resolution without a penalty to the sensitivity, and the
scheme validity does not suffer from nearby nuclear spins or relatively low coupling to the nuclear spins.

\emph{Sensing application ---} To evaluate the effectiveness of our scheme, we use it to determine
the molecular structure of a simple target molecule. To sense and address individual nuclear spins
in a crowded cluster in the molecule, the internuclear dipolar coupling needs to be suppressed which
we achieve by application of the WAHUHA \cite{waugh1968approach} decoupling sequences during the evolution.
This sequence averages the internuclear dipolar coupling to zero over the cycle time, at the cost of a
reduction factor in the dipolar coupling strength $1/\sqrt{3}$. Consider as an example the molecule malic
acid (Valine), which plays an important role in biochemistry and is not too large to prevent exact numerical
simulations. In Fig. \ref{spectra}c, we simulate the case of three $^{13}$C spins in this molecule, attached
to the diamond about 3.9 nm from the NV spin. The 1D NMR spectra show the information on the dipolar coupling
to the nuclear spins.  The dip height depends on the transverse coupling, as we have $S^d = 1 - \frac{1}{2}
\sin^2(pA_{wo}T/2)$ with $T$ the evolution time, and the longitudinal dipolar coupling can be derived from
the energy matching condition, which gives a pair of parameters $a_{\parallel_2}$ and $a_{\perp_2}$. Although
1D NMR spectra is not sufficient for determination of the position of the target without extra chemical information,
one can adjust the magnetic field orientation to determine precisely the 3D position of nuclear spins in the
molecule. Due to continuous collection of the signal, only a small number of measurements repetitions
are required to reach the SNR threshold. Assuming $C\approx  0.2$ is the combined contrast and photon
collection efficiency, and we ignore the time for initialization of our setup and the SWAP gates are ignored
here because of small values then, with $pA_{wo}\sim(2\pi)16$ Hz and $T=60$ ms, we estimate that the experimental
time for registering one dip in Fig. (\ref{spectra}c) is about 1.8 s, assuming that we take measurements
for $15$ frequency steps.

%Generalizing our method for heteronuclear sensing is possible with a small modification. Taking as an example the sensing of protons in a molecule on the diamond surface, the Larmor frequency of proton spins is about four times than that of $^{13}$C spin. One could apply corresponding rf fields to these two different nuclear spins, which will enable sensing by our method. Additionally, our method could be used for steady entanglement generation for quantum information processing (QIP) applications when $pA_{wo}\ll\Gamma^{eff}$.

\emph{Conclusion ---} We have designed a method that achieves the coupling between distant nuclear
spins mediated by a dissipatively decoupled electron spin of an NV center which suppresses the impact of
NV decoherence and relaxation processes on these nuclei. This enables high fidelity selective quantum gates
between two weakly coupled nuclear spins in a quantum register at ambient condition. Importantly for sensing
applications, our method enables the direct use of a nuclear spin as a quantum sensor rather than merely as
a memory. The controllable second-order coupling provides a tunable and sharp frequency filter for collecting
the signal continuously, hence increasing the SNR, while remaining insensitive to decoherence and relaxation of
the mediating NV. Therefore, making use of dissipation as a tool, the proposed NV-mediated scheme increases
the ability of detection and coherence control of nuclear spins as well as analysis of complex spin structures.

\emph{Acknowledgements ---} The authors thank L. McGuinness and Zhenyu Wang for helpful discussions and advice. This work was supported by the ERC Synergy grant BioQ, the EU projects DIADEMS, HYPERDIAMOND and EQUAM, the DFG CRC/TRR 21 and an Alexander von Humboldt Professorship.

\end{document}